\documentclass{ws-procs9x6}
\newif\ifpdf
\ifx\pdfoutput\undefined
\pdffalse 
\else
\pdfoutput=1 
\pdftrue
\fi

\newcommand{\ra}{\mbox{$\rightarrow$}}

\newcommand{\zpr}{\mbox{$Z'$}}
\newcommand{\mzp}{\mbox{$M_{Z'}$}}

\newcommand{\upr}{\mbox{$U(1)'$}}
\newcommand{\uprm}{\mbox{$U(1)'$}}

\newcommand{\skipblk}[1]{}                                                      
\def\bqa{\begin{eqnarray}}                                                      
\def\eqa{\end{eqnarray}}

\newcommand{\beq}{\[}                                             
\newcommand{\eeq}{\]}

\def\mxth{\mathsurround=0pt }
\def\xversim#1#2{\lower2.pt\vbox{\baselineskip0pt \lineskip-.5pt
  \ialign{$\mxth#1\hfil##\hfil$\crcr#2\crcr\sim\crcr}}}             
\def\simgr{\mathrel{\mathpalette\xversim >}}                                    
\def\simle{\mathrel{\mathpalette\xversim <}}                       
\begin{document}
 \ifpdf
\DeclareGraphicsExtensions{.jpg,.pdf,.mps,.png}
 \else
\DeclareGraphicsExtensions{.eps,.ps}
 \fi

\title{Beyond the MSSM\footnote{\uppercase{T}alk presented 
at {\it \uppercase{SUSY} 2003:
\uppercase{S}upersymmetry in the \uppercase{D}esert}\/, 
held at the \uppercase{U}niversity of \uppercase{A}rizona,
\uppercase{T}ucson, \uppercase{AZ}, \uppercase{J}une 5-10, 2003.
\uppercase{T}o appear in the \uppercase{P}roceedings.}}

\author{PAUL LANGACKER}

\address{Department of Physics and Astronomy\\
University of Pennsylvania\\ 
Philadelphia, PA 19104 USA\\ 
E-mail: pgl@hep.upenn.edu}

\maketitle

\abstracts{Specific quasi-realistic superstring constructions often predict the existence
of new TeV-scale physics which may be
different in character from bottom-up motivated constructions. I describe examples
of such beyond the MSSM physics, with special emphasis on heavy $ Z' $ gauge bosons and
associated exotic particles. Implications of an extra \uprm \ include
a viable scienario for electroweak baryogenesis,  a highly non-standard
light Higgs sector, and  \zpr-mediated contributions to rare $ B $ decays.
}

\section{Beyond the MSSM}


Even if supersymmetry holds,  the MSSM may not  be the full story.
The MSSM stabilizes but does not explain the hierarchy between the
electroweak and Planck scales, and it gives a plausible first step towards
the incorporation of quantum gravity. However, the MSSM does not address 
the other problems of standard model (SM). Moreover, the
$\mu$ problem (i.e., that the supersymmetric masses of the Higgs multiplets  and their
partners should be comparable to typical supersymmetry breaking scales) is introduced,
and some versions have difficulty in avoiding new flavor changing effects
and too large contributions to electric dipole moments.

One possibility is that the underlying new physics is at the Planck or a GUT scale,  
but even then there may be remnants surviving to the TeV scale. For example, specific string constructions often have extended gauge groups,
especially \uprm \ factors,  and associated exotic
chiral supermultiplets.
Such remnants might not
solve or be directly motivated by bottom-up solutions to the problems of the SM or MSSM,
but should instead be viewed as a consequence of the underlying high scale theory. For
this reason and others it is important to explore alternatives/extensions to MSSM.


\section{Possible new TeV-scale physics motivated by string constructions}
There have been a number of detailed investigations of concrete, semi-realistic 
superstring constructions, both heterotic (closed string) and interersecting brane (open 
string)~\cite{pascos}. Although it is unlikely that any fully realistic construction will
be found soon, this program  is useful for developing techniques and for suggesting new top-down
motivated physics that may survive to the TeV scale. Specific constructions 
have  led to such features as

\begin{itemize}
\item Direct compactification. The higher dimensional theory may break directly
to the MSSM in four dimensions (possibly with an extended gauge sector)
without going through a separate 4D GUT stage. This avoids the doublet-triplet problem
and the need for large representations for
GUT breaking and fermion/neutrino masses and mixings, that are hard to generate in some constructions.

\item The minimal gauge unification of the MSSM is often lost, either because a
non-standard unification is masked by
higher Ka\v c-Moody levels or exotic particle contributions to the runnings, or
because of   non-universal (moduli-dependent) boundary conditions.
The observed unification may be due to (accidental?) compensation of such effects,
or there may be new constructions in which these features are absent.

\item There are often new chiral supermultiplets, including exotic quarks
or leptons (i.e., with non-standard $SU(2) \times U(1)$   quantum numbers); extra
 Higgs doublets; standard model singlets; and fractionally charged particles.
 There is often mixing between lepton and Higgs doublets (i.e.,  $(\not{\! R}_P)$).
 	
\item In additional to the MSSM gauge group, there is often a quasi-hidden
gauge group. It is not truly hidden, because there are typically a few states
charged under the MSSM and hidden sectors. This may imply the existence
of fractional electric charges if the group is not asymptotically free (AF);
or charge confinement and light composites if it is AF~\cite{cls}.
There may also be gaugino condensation, leading to SUSY breaking and dilaton/moduli
	  stabilization~\cite{clw}.

\item The Yukawa sector is very model dependent, and there may
be different embeddings for the three families. Terms allowed by the
symmetries of the 4D effective field theory are often forbidden by string
selection rules, leading, e.g., to string-driven fermion textures.
Allowed terms are often totally off diagonal in the
fields, unlike many bottom-up phenomenological models for neutrino masses
or extended Higgs sectors. There may be relations between quark and lepton
Yukawas different than in simple GUTs.

\item There are often extended gauge sectors. Some of the extra factors may
be associated with a quasi-hidden sector. There are frequently additional (non-anomalous)
extra \upr \ factors which may survive to the TeV scale. Typically, the ordinary and
hidden-sector states are both charged under the \upr, and there are often a number of
SM singlets with \upr \ charges. The \upr \ couplings are
often family-nonuniversal, leading to flavor changing neutral currents (FCNC) after
flavor mixing~\cite{lp}.

\end{itemize}

\section{A heavy \zpr?}
Extra \upr \ gauge symmetries are predicted by many string constructions,
grand unified theories\footnote{However, GUTs  require extra
fine tuning for $\mzp \ll M_{\bf GUT}$ and may have problems with
proton decay.}, models of dynamical
symmetry breaking, and Little Higgs models. String constructions in
particular often predict extra \zpr s as well as numerous SM singlets that
are charged under the \upr.
Furthermore, radiative breaking of the electroweak symmetry 
(either in SUGRA or gauge mediated schemes)
often yields EW or TeV scale \zpr \ masses~\cite{ewscale} (unless the breaking is along a
flat direction, leading to breaking at an
intermediate scale~\cite{intermediate}).
The breaking may be due to negative mass$^2$ for a scalar $S$ 
charged under the \upr (driven by a large Yukawa coupling) or by an
$A$ term.

Limits on a possible \zpr \ mass and  the $Z-Z'$ mixing $\theta_{Z-Z'} $
from CDF~\cite{explim}
and precision experiments~\cite{indirect} are model dependent, but
typically $M_{Z'} > 500-800 $ GeV and
 $|\theta_{Z-Z'}| < {\rm  few} \times 10^{-3}$.
Discovery should be possible up to $\mzp \sim 5-8$ TeV at  the LHC or a linear collider, 
while diagnostics via asymmetries, $y$ distributions, associated production,
  and      rare decays should be possible  up to  to 1-2 TeV~\cite{cg}. 

\section{Implications of a \zpr}
A TeV scale \upr \ can lead to significant differences from the MSSM. In particular,
it leads to a solution to the $\mu$ problem~\cite{muprob1,muprob2} (i.e., why the
supersymmetric $\mu$ parameter in the superpotential,
$W_\mu = \mu H_u H_d$ is comparable to the electroweak (supersymmetry
breaking) scale). The \upr \ symmetry usually forbids
an elementary $\mu$ parameter, but allows a  superpotential term
$W = h S H_u H_d$, where $S$ is a SM singlet charged under the
\upr. The expectation value
$\langle S \rangle$
not only  breaks  the \uprm,  but also generates an effective
$\mu_{eff} = h \langle S \rangle$.
 This is similar to the Next to Minimal Supersymmetric Standard Model
 (NMSSM)~\cite{NMSSM},
  but avoids the NMSSM problems with cosmological domain walls~\cite{domain}.
  The \upr \ symmetry does not allow the superpotential term
 $W\sim \kappa S^3$, which is needed in the NMSSM, but its role is
 played by $D$ terms or superpotential terms involving several SM singlets
 (e.g., $\lambda S_1 S_2 S_3$). Many string constructions
 do not lead to terms like $S^3$, so the \upr \ models can be considered
 a string-motivated implementation of the NMSSM.
 
 Other implications of a \upr \ include: (a) the existence of new chiral supermultiplets
 with exotic SM quantum numbers, needed to cancel anomalies~\cite{erler}. These may
 be consistent with minimal gauge unification. (b) SM singlets charged under the
 \upr. (c) A non-standard sparticle spectrum~\cite{spectrum,ell}. (d)
 CP phase correlations~\cite{Demir}. (e) A TeV-scale \upr \ may forbid a
 large Majorana mass for the right-handed neutrino and therefore
 a conventional seesaw. Possibilities then include small Dirac masses
 (from higher-dimensional operators~\cite{hdo} or large extra dimensions),
 with implications for Big Bang Nucleosynthesis~\cite{bll} unless the
 breaking is such that the $\nu_R$ decouples~\cite{decoupling}; and
 TeV-scale seesaws. In this talk, I will comment more on
 (f) nonstandard Higgs/neutralino spectra~\cite{ell,higgs}, (g) electroweak 
 baryogenesis~\cite{ewbg},
 and (h) possible tree level FCNC effects relevant to rare $B$ decays~\cite{ll,fcnc}.

\section{A secluded sector model}
\label{secludedsector}
One possibility is for the \upr \ to be broken at the
 SUSY-breaking scale~\cite{muprob2}, by the same field $S$ which
 generates the effective $\mu$ parameter. One can have either:
 $\mzp \sim M_Z$,  if it is leptophobic (small leptonic couplings); or
 $\mzp \simgr 10 M_Z$ by modest tuning. Another possibility is to
 somewhat decouple the $Z'$ mass scale from  $\mu$ by
 allowing several SM singlets. 
In secluded sector models~\cite{ell}, the $Z'$ mass can be
 naturally large because it is associated with an
 approximately $F$ and $D$ flat direction, broken by a small $(\sim 0.05)$ Yukawa 
 coupling $\lambda$. In the examples in~\cite{ell}, there are 
four SM singlets, $S, S_{1,2,3},$ and two doublets $H_{1,2}$.
The superpotential is
\beq   W =
   h S H_1 H_2 + \lambda S_1 S_2 S_3, \eeq
where the first term is associated with $\mu$ and the second
with the approximate flat direction. 
The off-diagonal nature is motivated by string constructions.
The potential is then $V= V_F + V_D + V_{soft} $, where 
    \begin{eqnarray}
V_F &=& h^2 \left( |H_1|^2 |H_2|^2 + |S|^2 |H_1|^2 + |S|^2|H_2|^2\right)
\nonumber\\ &+&
\lambda^2 \left(|S_1|^2 |S_2|^2 + |S_2|^2 |S_3|^2 + |S_3|^2 |S_1|^2\right)
\, \nonumber
\end{eqnarray} 
 %
   \begin{eqnarray}
V_D &=& {{G^2}\over 8} \left(|H_2|^2 - |H_1|^2\right)^2 
\nonumber\\&+&
{1\over 2} g_{Z'}^2\left(Q_S |S|^2 + Q_{H_1} 
|H_1|^2 + Q_{H_2} |H_2|^2 + \sum_{i=1}^3 Q_{S_i}
|S_i|^2\right)^2 \, \nonumber
\end{eqnarray}  
{ where $G^2=g_1^{2} +g_2^2$,}
    and
  \begin{eqnarray}
V_{soft} &=& m_{H_1}^2 |H_1|^2 + m_{H_2}^2 |H_2|^2 + m_S^2 |S|^2 +
\sum_{i=1}^3 m_{S_i}^2 |S_i|^2
\nonumber\\&-&
(A_h h S H_1 H_2 + A_{\lambda} \lambda S_1 S_2 S_3 + {\rm H. C.})
\,  \nonumber \\
    &+& (m_{S S_1}^2 S S_1 + m_{S S_2}^2 S S_2 
+ m_{S_1 S_2}^2 S_1^{\dagger} S_2 + {\rm H. C.})
~\, \nonumber
    \label{softbreaking}
    \end{eqnarray} 
    For small $\lambda$ one finds    
 $\langle  S_i  \rangle \sim m_{S_i}/\lambda$, with the \upr \ breaking along 
 the $D$-flat direction $D(U(1)')\sim 0$, with smaller 
 $\langle  S  \rangle$ and $ \langle  H_i  \rangle$. Ensuring the correct minimum
 requires that the EW breaking is dominated by a large $ A_h h$ term
 rather than soft mass-squares, 
implying  $ \tan \beta \sim 1, \langle  S  \rangle\sim  \langle  H_i  \rangle$.
This leads to large doublet-singlet mixing in the Higgs and neutralino sectors,
and that the $(S, H_{u,d})$ and $S_i$ sectors are nearly decoupled. The
$m_{S S_i}^2$terms break two unwanted global $U(1)$ symmetries,
while the $m_{S_1 S_2}^2 $  term allows tree-level CP violation in the scalar
sector, with negligible contribution to electric dipole moments.

\section{Nonstandard Higgs and neutralino sector}
The extra \upr \ and the SM singlets in the secluded sector models
imply a rich Higgs and neutralino spectrum~\cite{ell,higgs}. In particular,
the tendency for $ \tan \beta \sim 1, \langle  S  \rangle\sim  \langle  H_i  \rangle$
for the model in Section \ref{secludedsector} leads to significant
 doublet-singlet mixing and therefore properties very different from the MSSM.
 That model involves nine neutralinos, with masses ranging from
 very light ($\simle 100$ GeV) to several TeV. Four are mainly in the
 secluded sector (\upr \ gaugino and $S_i$ Higgsinos), while five mainly
 overlap the SM sector (SM gauginos and $S, H_{u,d}$ Higgsinos).
 
\begin{figure}[htbp]
\begin{center}
\includegraphics[scale=.45]{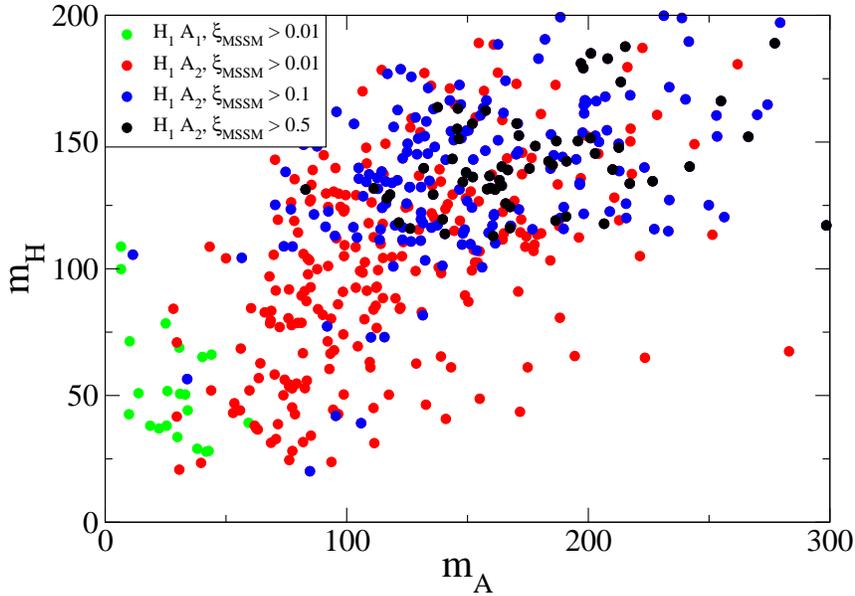}
\caption{Experimentally allowed scalar and 
pseudoscalar masses for typical parameters. $\xi_{\rm MSSM}$ is 
the overlap of the state with the Higgs doublets.}
\label{higgsmass}
\end{center}
\end{figure}

 Similarly, the neutral Higgs sector involves 6 scalars and 4 pseudoscalars
(which will mix with each other if one includes tree level CP breaking 
associated with the $m_{S_1 S_2}^2 $  term in
$V_{soft}$). These typically separate into two sectors, one mainly decoupled
(i.e., secluded). In principle, the lightest Higgs scalar can be considerably heavier
than in the MSSM because of new $F$ and $D$ terms,
\begin{eqnarray}
\label{mhlimit}
    M_h^2  & \le  & h^2 v^2 + (M_Z^2 - h^2 v^2) \cos^2{2 \beta}\nonumber \\
        &+&2 g_{Z^\prime}^2 v^2 (Q_{H_2} \cos^2 \beta^2 + \sin^2 \beta Q_{H_1})^2 
        + {3 \over 2} \frac{\cos^2 \beta  m_t^4}{v^2 \pi^2}
        \log {m_{\tilde{t_1}} m_{\tilde{t_2}} \over m_t^2},
        \nonumber
\end{eqnarray}
allowing masses  up to 185 GeV with all couplings perturbative
to $M_P$. Frequently, however, there are (nondecoupled) scalars and pseudoscalars
that are below the usual LEP SM and MSSM exclusion limits~\cite{lephiggs}
(or which are in parameter regions that are not theoretically possible in the MSSM).
These may have reduced couplings and
therefore be allowed experimentally because of the doublet-singlet mixing~\cite{higgs}.
Examples are shown in Figure \ref{higgsmass}.

\section{Electroweak baryogenesis}
Electroweak baryogenesis~\cite{RTWB} attempts to generate the
observed baryon to entropy ratio $n_B/s \sim 9 \times 10^{-11}$
by $B$-violating tunneling processes (sphalerons) at the time
of the electroweak phase transition (EWPT). In the ``off the wall''
scenario~\cite{CKNJPT}, nonequilibrium is provided the nucleation and
expansion of bubbles of true vacuum during  the EWPT.
The CP violation is due to the asymmetric reflection of chiral fermions from
the bubble wall back into the unbroken phase, where the sphaleron processes
convert the chirality into a baryon and lepton asymmetry (with $B-L$ conserved).
The transition must be strongly first order to quickly turn off the sphalerons.
Unfortunately, the SM is not strongly first order for the allowed Higgs masses,
and the CP violation from the CKM matrix is too weak. In the MSSM there are
new sources of CP violation, but the transition is not strong first order except
for a small parameter range involving a light Higgs ($<$ 120 GeV) and
$\tilde{t}$~\cite{RTWB,mssmbar}. The NMSSM allow a strong first order transition
for a large $A_h$ term $h A_h S H_1 H_2$~\cite{DFMHS}, but either suffers cosmological
domain wall problems or reintroduces the $\mu$ problem.

\begin{figure}[htbp]
\begin{center}
\includegraphics[scale=.6]{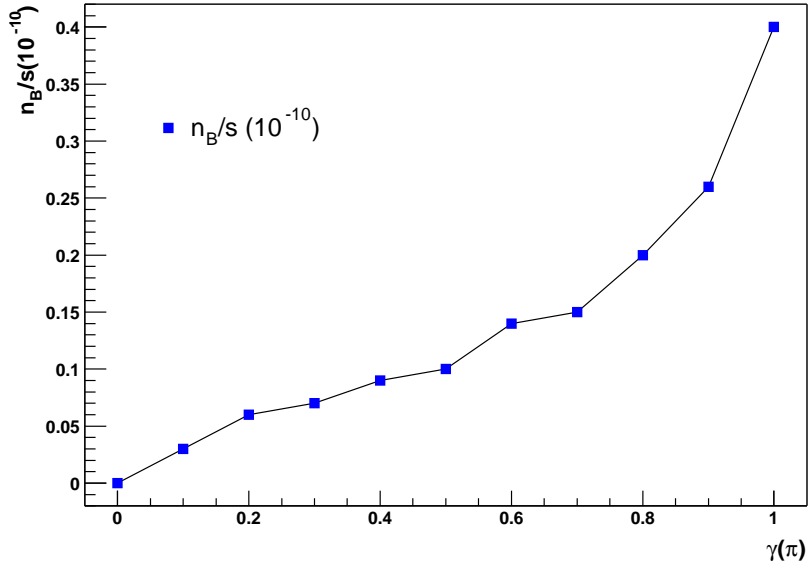}
\caption{Predicted baryon to entropy ratio as a function of a CP
violating phase $\gamma$ from the SM singlet sector for a 
particular set of input parameters. The asymmetry is
within a factor of 2 (i.e., within theoretical uncertainties)
 of the observed value for $\gamma$ close to $\pi$.}
\label{baryogenenesis}
\end{center}
\end{figure}
The secluded sector \upr \ model can easily account for the observed asymmetry within
theoretical uncertainties~\cite{ewbg}, even for a large $\tilde{t}$ mass.
 There are actually two transitions. The first
breaks \upr \ and the second $SU(2) \times U(1)$. The large $A_h$ term needed for the model
ensures that the second transition is strong first order. Finally, the tree-level CP
breaking allowed in the SM singlet sector allows the CP violation
in the bubble wall to be sufficiently large (most of the breaking is actually 
associated with the false (unbroken) vacuum)), without generating
significant new contributions to electric dipole moments.

\section{FCNC and rare $B$ decays}
The \upr \ couplings are often family-nonuniversal in string constructions,
leading to flavor changing neutral currents
(FCNC) mediated by the  \zpr \ (and the $Z$ from $Z-\zpr$ mixing)
after family mixing is turned on (because of GIM breaking). There may also
be FCNC due to mixing of the ordinary and exotic fermions.
 The strength depends on the mixing matrices $V _{\psi L}$ and $V _{\psi R}$,
 $\psi=u, d, e, \nu$, for the chiral fermions,
but only $V_{\rm \bf CKM} = V_{u_L} V_{d_L}^{\dagger}$ and
$V_{\rm \bf MNS} = V_{\nu_L} V_{e_L}^{\dagger}$ are known from experiment.
Stringent limits from  $K$ and $\mu$ decays imply that the  first
two families are universal~\cite{lp}.
However,  the third family could  have different couplings, leading, for example,
to effects in the forward-backward asymmetry for $Z$ decaying to $b \bar b$~\cite{indirect},
or to FCNC effects in rare $B$ decays~\cite{ll,fcnc}. The latter can be especially important
for rare decays which only occur at the loop level in the standard model~\cite{fcnc},
such as
 $B\ra \phi K$ or $  \eta' K$, for which there are possible anomalies. 
 They could also lead to an enhanced rate for $ B_s \ra \mu^+ \mu^-$
 or $ b \ra s \mu^+ \mu^-$
 (as an alternative to the MSSM with large $\tan \beta$), without
 signficantly modifying $b \ra s \gamma$.

\section{Acknowledgments}
It is a pleasure to thank my collaborators. This work was supported in part
by the U.S.~Department of
Energy under Grant No.~DOE-EY-76-02-3071. 


\end{document}